\documentstyle[12pt]{article}

\def\beqa{\begin{eqnarray}}
\def\eeqa{\end{eqnarray}}
\def\beq{\begin{equation}}
\def\eeq{\end{equation}}

\def\umunu{^{\mu\nu}}
\def\dmunu{_{\mu\nu}}

\def\ddemunu{_{;\mu\nu}}

\def\ddemu{_{;\mu}}  
\def\ddenu{_{;\nu}}  
\def\ddea{_{;\alpha}}  \def\udea{^{;\alpha}}

\def\bib#1{$^{\ref{#1}}$}

\let\alp=\alpha

\def\pr{{\it Phys. Rev.}\ }
\def\prl{{\it Phys. Rev. Lett.}\ }
\def\pl{{\it Phys. Lett.}\ }

\def\ijmp{{\it Int. Journ. Mod. Phys.}\ }

\def\cqg{{\it Class. Quantum Grav.}\ }

\def\f{F(\phi)}
\def\fp{F'(\phi)}
\def\fpp{F''(\phi)}
\def\p{\phi}
\def\pv{\varphi}
\def\v{V(\phi)}
\def\vp{V'(\phi)}

\begin{document}
\def\bib#1{[{\ref{#1}}]}
\begin{titlepage}
	 \title{ Asymptotic Freedom from Induced Gravity Cosmology}

 \author{{S. Capozziello$^{1,3}$, R. de Ritis$^{2,3}$}
\\ {\em $^{1}$Osservatorio Astronomico di Capodimonte, Via Moiariello 16}
 \\{\em I-80131 Napoli, Italy}
\\ {\em $^{2}$Dipartimento di Scienze Fisiche, Universit\`{a} di Napoli,}
\\ {\em $^{3}$Istituto Nazionale di Fisica Nucleare, Sezione di Napoli,}\\
   {\em Mostra d'Oltremare pad. 19 I-80125 Napoli, Italy.}}
	      \date{}
	      \maketitle
	      \begin{abstract}
We give  conditions to obtain cosmological
 asymptotic freedom 
in   scalar--tensor theories of gravity.
We show that this
feature can be  achieved in  FRW flat
spacetimes since we obtain singularity free solutions  where the effective
gravitational constant $G_{eff}\rightarrow 0$ for $t\rightarrow -\infty$
and, for some of them, $G_{eff}\rightarrow G_{N}$ for $t\rightarrow\infty$,
where $G_{N}$ is the Newton constant.

	      \end{abstract}

\vspace{20. mm}
PACS: 98.80 Dr\\
e-mail:\\
 capozziello@astrna.na.astro.it\\
deritis@axpna1.na.infn.it
	      \vfill
	      \end{titlepage}
\section{\normalsize \bf Introduction}
Nonminimal coupling between gravity and one (or more than one) scalar
field is recently become a general "paradigm" with respect to formulate
any effective theory of matter and gravity which tries
to take into account cosmology and particle physics \bib{birrell}. 
We have to define
an effective gravitational constant $G_{eff}$ and an effective cosmological
constant $\Lambda_{eff}$ which we need for renormalization
(at least at one--loop level)  and that  have
to furnish the present observed values $G_{eff}\rightarrow G_{N}$,
$\Lambda_{eff}\rightarrow 0$.
Furthermore, nonminimally
coupled theories of gravity  furnish cosmological models which
bring to an end the inflationary stage without any imposed fine--tuning:
 in other words, the
shortcomings of original inflationary models (the so called {\it old} 
and {\it new} inflation \bib{guth}) are naturally avoided
by supposing a variation of Newton constant which regulates the phase
transition from the false--vacuum state to the true--vacuum state
\bib{la}.
In this scheme,  gravity is an {\it induced} interaction 
which could result from an average effect of the other fundamental forces.
Then, we have to search for similar features between gravity and the other
interactions. 
As it is
shown in \bib{matter}, it is possible to relate the today
observed Newton constant with the self coupling constant $\lambda$ of an
effective
scalar field potential. This constant is related to the gauge coupling
$\alp_{GUT}$ being $\alp_{GUT}=g^{2}/4\pi$, $\lambda=g^{2}$
and $G_{N}\propto\lambda$
\bib{quigg}. 

In particular, we know that any
force mediated by the exchange of non--Abelian gauge quanta has a property 
called "asymptotic freedom", 
which means that the effective strength of 
the interaction tends to zero at short distances, or, in other words,
if the energy of the system diverges.
The existence of asymptotic freedom of strong interactions was indicated
by a series of experiments  which are the high energy
counterparts of the Rutherford experiments with the alpha particles
\bib{meyerhof}. 
This scheme was applied also to 
high energy electrons which were scattered on proton 
targets \bib{appelquist}.
The goal was to study the
internal structure of proton (deep--inelastic scattering). 
The result was that when the exchanged energy--momentum became larger,
the interaction among the quarks in the proton became weaker
\bib{bjorken}.
In principle, we can seek similar behaviours also in gravitational 
interaction
\bib{markov},
 but the lack of full quantization
does not allow us to apply the quantum interpretative scheme 
which holds in QCD.
As a matter of fact, gravity is a "classical" theory since we have not yet
a quantum interpretation of spacetime. However, if we adopt an induced
gravity interpretation, we can assume that the average effects
of the other material interactions lead, in some sense,
the gravitational interaction, as discussed above.
 These effects have to mimics
strong interactions at high energies (i.e. at short distances) but have to
make one recover Einstein standard gravity at low energy limits.
From a cosmological point of view, "short distances" mean "early times"
and  "divergence of matter--energy density",
while "low energy limits" mean that the cosmic time $t\rightarrow\infty$.
In other words, we shall have a sort of "gravitational asymptotic freedom"
if
$\lim_{t\rightarrow -\infty}G_{eff}\longrightarrow 0,$
and
 recover the standard gravity if
$
\lim_{t\rightarrow \infty}G_{eff}\longrightarrow G_{N}\,.
$
However, we are supposing to have singularity free cosmological
solutions for which 
the scale factor of the universe $a(t)$ is defined on the interval
$-\infty <t< \infty$).

In this paper, we address  the issue to find gravitational 
 asymptotic freedom
 in a nonminimally coupled theory.
In Sec.2, we give the conditions of how asymptotic freedom 
 can be found  in a cosmological context. 
In Sec.3, we discuss cosmological models
 without singularity, by which it is possible
 to find the property of asymptotic freedom
 when cosmic time  $t\rightarrow -\infty$.
Furthermore, some of them allow to recover standard Einstein gravity
at present. In Sec.4, we draw conclusions.

\section{\normalsize\bf Conditions for Cosmological Asymptotic
Freedom}

We start our analysis to search for gravitational asymptotic freedom
from a generic action where a scalar field $\p$ is nonminimally coupled
with gravity:
\beq
\label{3}
{\cal A}=\int d^{4}x\sqrt{-g}\left[\f R +\frac{1}{2}
g\umunu \phi\ddemu \phi\ddenu-V(\phi)\right]\,.
\eeq
Here $\v$ is the potential and $\f$ the coupling for the field $\p$.
 We
adopt the units $8\pi G_{N}=\hbar=c=1$.

The field equations are obtained by a variation with respect to $g\dmunu$
\beq
\label{4}
\f G\dmunu=-\frac{1}{2}T\dmunu-g\dmunu\Box\f+\f\ddemunu,
\eeq
where $G\dmunu$ is the Einstein tensor and
$
T\dmunu=\phi\ddemu \phi\ddenu-\frac{1}{2} g\dmunu \phi\ddea \phi\udea+
g\dmunu\v,
$
is the "bare"  energy-momentum tensor of the scalar field.
By a  variation with respect to $\phi$, we obtain the generalized
Klein-Gordon equation:
\beq
\label{6}
\Box \p-R\fp+\vp=0,
\eeq
where the prime indicates  the derivative with respect to $\p$.

Let us consider now a homogeneous and isotropic
 cosmology 
where  $a=a(t)$ is the scale factor of the universe and the field
$\p=\p(t)$ is a function  of cosmic time only.
The Lagrangian density in (\ref{3}) becomes
\beq
\label{7}
{\cal L}=6a\dot{a}^{2}\f+6\dot{a}\dot{\p}a^{2}\fp
-6ka\f +
a^{3}\left[\frac{1}{2}\dot{\p}^{2}-\v \right]\,,
\eeq
where the dot is the time derivative and $k$ is the spatial 
curvature constant.
From now on, we assume $k=0$.
The Einstein equations (\ref{4}) become
\beq
\label{8}
H^{2}= -\frac{1}{\f}\left[H\dot{\p}\fp +\frac{1}{12}\dot{\p}^2
+\frac{1}{6}\v \right]\,,
\eeq
\beq
\label{9}
\dot{H}+H^{2}= - \frac{1}{2 \f}\left[H\dot{\p}\fp - \frac{1}{3}\dot{\p}^2
+\frac{1}{3}\v +\ddot{\p}\fp + \dot{\p}^2\fpp \right]\,,
\eeq
 and the Klein--Gordon equation is
\beq
\label{klein}
\ddot{\p}+6\left(\dot{H}+2H^{2}\right)\fp+3H\dot{\p}+\vp=0\,,
\eeq
where $\fp\equiv d\f/d\p$ and $H=\dot{a}/a$ is the Hubble parameter.
We have to note that the effective gravitational
constant is
\beq
\label{effective}
G_{eff}\equiv\frac{1}{2\f}\,,
\eeq
and, for $\f=-1/2$,
we recover the standard Einstein equations of Friedman
cosmology.

We rewrite Eqs.(\ref{8})   as:
\beq
\label{10}
H^{2}+H\frac{\dot{F}(\p)}{\f}+\frac{\rho_{\p}}{6\f}=0\,,
\;\;\hbox{where}\;\;
\rho_{\p}=\frac{1}{2}\dot{\p}^{2}+\v\,;
\eeq
$\rho_{\p}$ is the matter--energy density  
associated to the (minimally coupled)
scalar field; it can be considered a sort of "bare" energy density.

As we  discussed in \bib{quartic}, from Eq.(\ref{4}), we can
say that, in general, in these nonminimally coupled gravity theories there
is no singularity. As consequence, in such cosmologies we
have that the physical quantities are all defined
in the time interval  $(-\infty,+\infty)$.

Anyway, we have to keep in mind the question of whether or not the theory
admits an absolutely stable ground state (at classical level) in which
$\phi$ assumes a constant value and the Ricci tensor $R\dmunu$ is 
proportional to $g\dmunu$, i.e. spacetime is either Minkowski, de Sitter,
or anti--de Sitter. In curved spacetimes (that is if nonminimal coupling
terms appears) the energy density is not a good 
criterion for the stability so that conditions on the parameters in $\f$
and $\v$ have to be imposed. In general, the stability request 
and the recovering of standard gravity restrict
the range of possible values of parameters like $\xi$, $\Lambda$, $m^{2}$
and $\lambda$ in theories where $\f=1-\xi\phi^{2}$ and
$\v=\Lambda+m^{2}\phi^{2}+\lambda\phi^{4}$ as discussed in \bib{deser}.
In what follows, we have to take into
account such issues since asymptotic states have to be stable. In the
examples of next section,  the ranges of parameters are specified.

In general, what  is needed for recovering the cosmological 
asymptotic freedom
is that, when $t\rightarrow-\infty$,
\beq
\label{014}
\rho_{\p}=\rho_{\p}\left(\dot{\p}(t),V(\p(t))\right)\longrightarrow
\infty\,,\;\;\;
a(t)\longrightarrow 0\,,
\eeq
which implies
\beq
\label{016}
F(\p(t))\longrightarrow\infty\,,
\eeq

The  condition on $\rho_{\p}$ tell us that the "bare" energy density 
has to
diverge in order to follow the analogy with the elementary particle case.
The condition on $a(t)$ takes into account that, differently from high
energy physics, in cosmology any length varies in connection with the 
dynamical behaviour of spacetime. That condition then tells us that any
lengths  approach to zero. 
Before a rigorous analysis, first we will discuss such problem in a 
qualitative way. So doing, we will quickly understand some aspects 
concerning the coupling $\f$, the potential $\v$, and the time behaviour of 
$a(t)$ and $\p(t)$. Let us assume that we obtain conditions
(\ref{014})--(\ref{016}) using exponential functions, that is
 \beq
 \label{14'}
 \rho_{\p}(t)\sim\rho_{0}\exp(-k_{2}t)\,,\;\;
 a(t)\sim a_{0}\exp(\Lambda t)\,,\;\;
 F(\p(t))\sim F_{0}\exp(-k_{1}t)\,,
 \eeq
 where $k_{1,2}$, $\Lambda$ are positive constants.
Actually, in this qualitative analysis, we are assuming that
$G_{eff}\rightarrow 0$ for $t\ll 0$.
From Eq.(\ref{10}), we get:
\beq
\label{casimiro}
\frac{\rho_{\p}}{6F}=\frac{\rho_{0}}{6F_{0}}e^{({k_{1}}-{k_{2}})t}=
\Lambda (k_{1}-\Lambda)\,,
\eeq
and we see that it must be $k_{1}\geq k_{2}$. 
We have to discuss two cases:
\vspace{3. mm}\\
 $i)$ $\Lambda=k_{1}$
in the case  $k_{1}>k_{2}$;
\vspace{3. mm}\\
 $ii)$ 
$\Lambda^{2}-\Lambda k_{1}+(\rho_{0}/F_{0})=0$ if $k_{1}=k_{2}$. In any 
case,
we find that 
\beq
\label{paperoga}
\frac{\rho_{\p}}{6F}=\hbox{const}=k_{3}\,,
\eeq
where $k_{3}=0$ for $k_{1}>k_{2}$ or $k_{3}=\Lambda(k_{1}-\Lambda)$ 
for $k_{1}=k_{2}$.
From Eq.(\ref{9}), we find
\beq
\Lambda^{2}=\frac{1}{2}\Lambda k_{1}-\frac{1}{6}\frac{\dot{\p}^{2}}{F}
-\frac{V}{6F}-\frac{k_{1}^{2}}{2}\,.
\eeq
By using (\ref{casimiro}), we get
\beq
\label{18}
\frac{\v}{6\f}=\sigma_{0}\,,
\eeq
where $\sigma_{0}$ is a constant. We have some different cases:\\
$a)$ $k_{1}>k_{2}$,
in this case we get
$\sigma_{0}=-\frac{\dot{\p}^{2}}{12F}=-\Lambda^{2}.$\\
$b)$ $k_{1}=k_{2}$,
here we have $\sigma_{0}=\frac{\rho_{0}}{6F_{0}}-\frac{k_{1}}{2}
(1-3\Lambda).$\\
$c)$ Choosing from the very beginning $\v=\lambda$, we get from 
(\ref{016}) that (asymptotically) $\sigma_{0}=0$. Of course, we can also
choose $\lambda=0$, then (constantly) is $\sigma_{0}=0$.
In both cases $a)$, $b)$ we get that $\p(t)\sim \exp(-k_{2}t/2)$, 
that is
$\f\sim \p^{2}$ and then $\v\sim \p^{2}$
(we are not taking into any consideration the role of the constants 
appearing in $\sigma_{0}$ in those two different cases). The two
cases $c)$ are respectively the cosmological constant case in presence
of nonminimal coupling, and the Brans--Dicke type model (with no potential).

We will now discuss in a rigorous way the above situations and so doing
we will deduce that $F(\p(t))\rightarrow\infty$ (i.e. $G_{eff}
\rightarrow 0$) for $t\ll 0$ in an exponential way.

\section{\normalsize \bf Cosmological models with asymptotic freedom}

Now we discuss models where the above 
hypotheses hold. We have to note that the 
gravitational asymptotic freedom can  depend or not on the initial
conditions, this conditioned on the value of $\sigma_{0}$. In fact, as we 
shall see below, we have models where asymptotic freedom holds for
general solutions (in the sense that initial conditions have not to be 
specified) and models where it holds for a restricted range of initial data.
This feature depends on the ratio of $\v$ and $\f$, that is on $\sigma_{0}$.

\subsection{\normalsize \bf The case with $\v=\lambda\p^{2}$}
This is the case of the so called "free" effective potential coming from
the one--loop approximation of a scalar field.
The method to seek the general solutions (we have called it the
{\it N\"other Symmetry Approach}) is discussed in \bib{nmc}. 
In that case, the coupling and the potential  have the forms
\beq
\label{28}
\f=k_{0}\p^{2},\;\;\;\;\;\;\;\v=\lambda\p^{2}\,,
\eeq
where $k_{0}$ and $\lambda$ are free parameters, and, furthermore, it has to be
 $k_{0} \neq 1/12$ in order to 
avoid the degeneration (i.e the $\p$--part of Hessian determinant is zero) 
of the 
Lagrangian (\ref{7}).
Solving exactly the system (\ref{8})--(\ref{klein}) we find that
the scale factor and the scalar field evolve as
\beqa
a(t)&=&\left[
c_{1}e^{{\Lambda_{0}} t}+c_{2}e^{-{\Lambda_{0}} t}\right]\times \nonumber\\
    & & \exp\left\{ -\frac{2}{3}\left[
c_{3}\arctan \sqrt{\frac{c_{1}}{c_{2}}} e^{{\Lambda_{0}} t}
+c_{4}\ln (c_{1}e^{{\Lambda_{0}} t}+c_{2}e^{-{\Lambda_{0}} t})\right]\right\},
\label{29}
\eeqa
and
\beq
\label{30}
\p (t)=\frac{
\exp\left[
c_{3}\arctan \sqrt{\frac{c_{1}}{c_{2}}} e^{{\Lambda_{0}} t}
+c_{4}\ln (c_{1}e^{{\Lambda_{0}} t}+c_{2}e^{-{\Lambda_{0}} t})
\right]}{c_{1}e^{{\Lambda_{0}} t}+c_{2}e^{-{\Lambda_{0}} t}}.
\eeq
We see, from (\ref{29}), that the asymptotic behaviour of $a(t)$
(for $t\rightarrow \pm\infty$) is de Sitter like.
Then the Hubble parameter is
\beq
\label{31}
H=\Lambda_{0} \left(1-\frac{2}{3}\frac{\xi_{1}}{\xi_{2}}\right)
\left(\frac{c_{1}e^{{\Lambda_{0}} t}-c_{2}e^{-{\Lambda_{0}} t}}{c_{1}e^{{\Lambda_{0}} t}
+c_{2}e^{-{\Lambda_{0}} t}}\right)-
\left(\frac{c_{3}}{c_{1}e^{{\Lambda_{0}} t}
+c_{2}e^{-{\Lambda_{0}}t}}\right)\,,
\eeq
where
\beq
\label{32}
\Lambda_{0}=\sqrt{\frac{2\lambda \xi_{2}}{\xi_{1}(\xi_{1}-\xi_{2})}},
\;\;
c_{3}=\frac{{\cal F}_{0}\sqrt{c_{1}c_{2}}}{\xi_{2}\Lambda_{0}},
\;\;c_{4}=\frac{\xi_{1}}{\xi_{2}}\,,\;\;
\xi_{1}=1-12k_{0}\,,\;\;\xi_{2}=1-\frac{32}{3}k_{0}\,.
\eeq
The constants $c_{1}$, $c_{2}$, $c_{3}$  are  the initial data 
and ${\cal F}_{0}$
is a constant of motion \bib{nmc}. The asymptotic behaviour of these solutions
are 
\beq
\label{33}
\lim_{t\rightarrow -\infty}a(t)=
a_{0}\exp
\left[
-\sqrt{\frac{\lambda(1-8k_{0})^{2}}{2k_{0}(12k_{0}-1)(3-32k_{0})}}t\right]\,,
\eeq
\beq
\label{34}
\lim_{t\rightarrow -\infty}\p(t)=
\p_{0}\exp
\left[
-\sqrt{\frac{8\lambda k_{0}}{(12k_{0}-1)(3-32k_{0})}}t\right]\,.
\eeq
Coherently with (\ref{29}), we see that $a(t)$ diverges for 
$t\rightarrow -\infty$, that is it has to be a de Sitter behaviour.
From (\ref{33}),
we immediately recover asymptotic freedom
($G_{eff}\rightarrow 0$) for $t\ll 0$ (actually, we know the complete 
integral of the model and
we recover the same  result for $t\gg 0$).
Furthermore, we have to say that (\ref{18}) is not a  condition to obtain
asymptotic freedom in this case, since 
$\v/(6\f)=\sigma_{0}=\lambda/6k_{0}$ always holds.

\subsection{\normalsize \bf The string--dilaton cosmology case}

A string--dilaton four--dimensional effective action, neglecting the torsion
terms
and  other scalar fields except the dilaton $\pv$, is \bib{stringhe}
\beq
\label{01}
{\cal A}=\int d^{4}x\sqrt{-g} e^{-2\pv}\left \{\frac{1}{2}[ R +4
g\umunu  \pv\ddemu \pv\ddenu-2\Lambda]\right \} \;.
\eeq
This action is nothing else but a particular case of the most
 general action (\ref{3}),
when we take the positions
\beq
\label{03}
\p=2e^{-\pv},\;\;\;F(\p)=\frac{1}{8}\p^{2}=\frac{1}{2}e^{-2\pv},\;\;\;
\v=e^{-2\pv}\Lambda\,,
\eeq
(we see that $\f\sim\p^{2}$, $\v\sim\p^{2}$ and $\v/(6\f)=\sigma_{0}$).

In a FRW flat metric,  the action (\ref{01})
gives rise to a Lagrangian density:
\beq
\label{08}
{\cal L}=e^{-2\pv}[3\dot{a}^{2}a-6\dot{a}a^{2}\dot{\pv}+2a^{3}\dot{\pv}^{2}-
a^{3}\Lambda]\,.
\eeq
whose equations of motion,
 by the transformations (\ref{03}), can be recast in the form of
system (\ref{8})--(\ref{klein}).
Also here, condition (\ref{18}) holds 
at any time (with $\sigma_{0}\neq 0$).
The general solution of the dynamics
is (using also here the N\"other Symmetry Approach)
\beq
\label{033}
a(t)=a_{0}\exp\left\{\mp \frac{1}{\sqrt{6}} \arctan \left[\frac{
1-2e^{4\lambda \tau}}{2e^{2\lambda \tau}\sqrt{1-e^{4\lambda \tau}}}\right]
\right\},
\eeq
\beq
\label{034}
\pv(t)= \frac{1}{4} \ln\left[ \frac{2\lambda^{2}e^{4\lambda \tau}}
{\left( 1-e^{4\lambda \tau}\right)}\right]
\mp \frac{1}{\sqrt{6}} \arctan \left[\frac{
1-2e^{4\lambda \tau}}{2e^{2\lambda \tau}\sqrt{1-e^{4\lambda \tau}}}\right]
+\pv_{0},
\eeq
where $\tau =\pm t$, $\lambda^{2} =\Lambda /2$. In this solution the "scale 
factor duality" (i.e. the property that if $a(t)$ is a solution
$a(t)^{-1}$ is a solution too) is evident \bib{stringhe}. 
Using (\ref{033}) and (\ref{034}) at $t\ll 0$, 
the asymptotic freedom is  easily recovered.

\subsection{\normalsize \bf The case with $\v=\Lambda$}
This is a very interesting case since the solutions allow to recover
 the standard Einstein gravity 
at $t\rightarrow\infty$,
and the asymptotic freedom at $t\rightarrow-\infty$.
 
Also for $\v=\Lambda$, 
the dynamical equations (\ref{8})--(\ref{klein})
 can be exactly solved using the N\"other Symmetry Approach \bib{nohair}.
The existence of the N\"oether symmetry
selects a coupling of the form
\beq
\label{36}
\f = \frac{1}{12}\p^{2} + F_{0}'\p +F_{0}\,,
\eeq
where $F_{0}'$ and $F_{0}$ are integration constants. 
The general solution of the system (\ref{8})--(\ref{klein}) is \bib{nohair}
\beq
\label{39}
a(t)=\left[c_{1}e^{\lambda t}
+c_{2}e^{-\lambda t}\right]^{1/2}\,,\;\;
\p(t)=\frac{{\cal J_{0}}}{\sqrt{c_{1}e^{\lambda t}
+c_{2}e^{-\lambda t}}}{\cal K}+
\frac{c_{3}}{\sqrt{c_{1}e^{\lambda t}+c_{2}e^{-\lambda t}}}-6F_{0}'\,,
\eeq
where $c_{1}$, $c_{2}$ and $c_{3}$ are integration constants and
$\lambda =\sqrt{-2\Lambda/3\bar{\cal H}}$ 
with ${\bar{\cal H}}=F_{0}-3{F_{0}'}^{2}$ (the $\p$--part of the
Hessian determinant of
${\cal L}$).
${\cal J_{0}}$ is a constant of motion \bib{nohair} and 
\beq
{\cal K}=\int \frac{dt}{\sqrt{c_{1}e^{\lambda t}+c_{2}e^{-\lambda t}}}\;,
\eeq
is an elliptical integral of first kind.
It is easy to see that the asymptotic freedom 
(that is $F_{-\infty}\rightarrow\infty$) is recovered by choosing  the
initial condition $c_{2}=0$, otherwise we have 
$F_{\pm\infty}\rightarrow$ const.
In other words, we recover always standard gravity for $t\rightarrow\infty$,
being $G_{eff}\rightarrow$ const=$G_{N}$, but asymptotic freedom is
recovered only for a certain set of initial conditions. Furthermore, we
have to note that, for $c_{2}=0$, this model allows to recover both
standard gravity and asymptotic freedom with the "same" cosmological
de Sitter behaviour. Finally, referring to the discussion in Sec.2, this
case corresponds to get asymptotically $\sigma_{0}=0$.

\subsection{\normalsize \bf The Brans--Dicke case}
A pure Brans--Dicke action, without ordinary matter contributions, 
 can be written as
\beq
\label{23}
{\cal A}=\int d^{4}x\sqrt{-g}\left[{\bf \p}R+
\frac{\omega({\bf \p})}{{\bf \p}}g\umunu{\bf \p}\ddemu{\bf \p}\ddenu\right]\;.
\eeq
It can be recast in a usual nonminimally coupled form, like (\ref{3}),
with $\v=0$ by the transformations \bib{nohair}
\beq
\label{24}
{\bf \p}=\f\;,\;\;\;\;\;\;\;\omega({\bf \p})=\frac{\f}{2\fp}\;,
\eeq
(where we are not specifing the function $\f$).
In a homogeneous and isotropic metric, we recover 
the Lagrangian (\ref{7}) and the equations
(\ref{8})--(\ref{klein}) (in which we take $\v=0$ or $\sigma_{0}=0$ 
constantly, as in case $c)$ with $k_{1}=k_{2}$).
By imposing into (\ref{8})--(\ref{klein})
\beq
\label{25}
a(t)=a_{0}e^{\Lambda t}\,,
\eeq
we get
\beq
\label{26}
\p(t)\sim e^{\frac{-3\Lambda}{2}t}+\p_{0}\sim e^{\frac{-3\Lambda}{2}t}
\,,\;\;
F(\p(t))\sim c_{2}e^{-3\Lambda t}\,,
\eeq
we see that also here, asymptotically, $\f\sim\p^{2}$.
Actually, the presence of the constants $(c_{1}, c_{2}, \p_{0})$
tells us that $\f$ has to be a more complicated function of
$\p$. Even if we do not have the complete control of the time 
evolution of the model, we have shown that it has the feature
of asymptotyc freedom. We have to note that (\ref{25}) is only
a particular solution of the system; for a discussion of the
definition of $\Lambda$ in this case without potential see
\bib{nmc}.

\section{\normalsize \bf  Discussion and Conclusions}
In our first qualitative considerations, we have used 
exponential asymptotical 
behaviour for $a(t)$ and $\rho_{\p}(t)$, getting $H=\dot{a}/a$ constant.
Let us approach the issue to get asymptotic freedom from a more general
point of view.

Eq.(\ref{10}) can be rewritten in an integral form as:
\beq
\label{12}
F(\p(t))=\frac{\tilde{F}_{0}}{a}\exp\int -\left[\frac{\rho_{\p}}{6HF}\right]dt .
\eeq
Let us now assume that asymptotically (i.e. for $t\rightarrow-\infty$),
\beq
\label{13}
\frac{\rho_{\p}}{6H\f}=\Sigma_{0}\,,
\eeq
where $\Sigma_{0}$ is a positive constant.
Then we have
\beq
\label{14}
F(\p(t))=\frac{\tilde{F}_{0}}{a}\exp\left(-\Sigma_{0}t\right)\,.
\eeq
Hypothesis (\ref{13}) is more general with respect to what we
have realized untill now using exponential asymptotic funtions
for $a(t)$ and $\rho_{\p}(t)$. In fact, we are now assuming that $H$ is not,
a priori, a constant. Of course, if we assume, or if we show that $H$ is 
constant, we get that $\rho_{\p}/(6\f)$ is a constant too, that is
we restore (\ref{paperoga}).
Hypothesis (\ref{13}), being a relation among $(a,\dot{a},\p,\dot{\p})$,
has to be compatible with the Klein--Gordon equation (\ref{klein}), then
we get
\beq
\label{15}
6\dot{H}F\Sigma_{0}+6H\dot{F}\Sigma_{0}+3H\dot{\p}^{2}+6\dot{H}\dot{F}+
12H^{2}\dot{F}=0\,.
\eeq
Eq.(\ref{9}), by Eq.(\ref{8}), can be recast in the form
$\dot{\p}^{2}=4F\dot{H}-2H\dot{F}+2\ddot{F}.$
With a little algebra,
we obtain
\beq
\label{17}
\dot{H}+2H^{2}-3\Sigma_{0}H+\Sigma_{0}^{2}=-\frac{\v}{\f}\,.
\eeq
Let us now suppose that in the above limit $(t\rightarrow -\infty)$ 
the condition (\ref{18}) holds.
 Eq.(\ref{17}) becomes 
\beq
\label{19}
\dot{H}+2H^{2}-3\Sigma_{0}H+\Sigma^{2}_{0}+6\sigma_{0}=0\,,
\eeq
which is exactly solvable. 
The solution of this (asymptotic) equation  is
\beq
\label{20}
H=\frac{\lambda_{1}Ce^{{\lambda_{1}}t}+
\lambda_{2}e^{{\lambda_{2}}t}}{2\left[Ce^{{\lambda_{1}}t}+
e^{{\lambda_{2}}t}\right]}\,,
\eeq
where $C$ is the integration constant and
\beq
\label{21}
\lambda_{1,2}=\frac{3}{2}\Sigma_{0}
\pm\frac{1}{2}\sqrt{\Sigma_{0}^{2}-48\sigma_{0}}\;.
\eeq
It is worthwhile to note that, asymptotically for $t\rightarrow -\infty$, 
$H$ converge to a constant then de Sitter behaviour is recovered
(it is important to stress that in this way we get (\ref{paperoga})
under the hypothesis $\v/(6\f)=$ constant as is clear from (\ref{17})).

Being $H=\dot{a}/a$, we get from (\ref{20}) the scale factor
of the universe
\beq
\label{22}
a(t)=a_{0}\sqrt{Ce^{\lambda_{1}t}+e^{\lambda_{2}t}}
\;.
\eeq
whose asymptotic behaviour strictly depends on the signs and the values of
$\Sigma_{0}$ and $\sigma_{0}$.
Inserting (\ref{22}) into (\ref{14}), we get
\beq
\label{freedom}
F(\p(t))=\left(\frac{\tilde{F}_{0}}{a_{0}}\right)
\frac{e^{-\frac{7}{4}\Sigma_{0}t}}{\sqrt{Ce^{\frac{\mu}{2}t}+
e^{-\frac{\mu}{2}t}}}\,,
\eeq
where
$\mu=\sqrt{\Sigma_{0}^{2}-48\sigma_{0}},$
is a positive definite constant (it has to be $\Sigma_{0}\geq 48\sigma_{0}$
since $H$ is a real number).
Eq.(\ref{freedom}) has to diverge for $t\ll 0$ to get asymptotic
freedom. This situation, which is always true,  is, in any case, 
compatible with the reality condition
then we always get
\beq
\label{pluto}
F(\p(t))\sim e^{-\gamma t}\,,\;\;\;\;\;\;\;\;\hbox{for}\;\;\;\;t\ll 0\,,
\eeq
where $\gamma=\gamma(\Sigma_{0},\sigma_{0})$ is a constant determined by
a $\Sigma_{0}$ and $\sigma_{0}$. On the other side, the scale factor of the
universe converges exponentially to zero for
any combination of $\Sigma_{0}$ and $\sigma_{0}$, but diverges as 
$a(t)\sim e^{-\mu t/4}$ when
$\Sigma_{0}>0$, $\sigma_{0}<0$ and $\Sigma_{0}<6|\sigma_{0}|$
(the behaviour (\ref{pluto}) is not altered by this last condition).
It is interesting to note that in both cases (that is when $a(t)\rightarrow 0$
and $a(t)\rightarrow\infty$ for $t\ll 0$) we loose the gravitational
interaction; in other words, if a given length converges or diverges
the result is the same: 
the first situation can be seen as an analog of QCD, the second one
as the lack of interaction due to the fact that test particles are
brought to infinite distance.
We see that also using the more general hypothesis (\ref{13}), 
the dynamics leads again to exponential functions for $a(t)$ and
$\p(t)$ as well as to a nonminimal coupling which, in general, is
still $\f\sim\p^{2}$ as we can easily obtain  putting the above results
into (\ref{klein}). Of course such behaviours are controlled by the two
parameters $\Sigma_{0}$, $\sigma_{0}$.

As it emerges from these last considerations, it is clear that 
Eqs.(\ref{15})--(\ref{19}) are the asymptotic form of the system of
equation (\ref{8})--(\ref{klein}): however, solving 
the system (\ref{15})--(\ref{19}) does not mean that these
"asymptotic" solutions are the asymptotic behaviour of the solutions 
of the system (\ref{8})--(\ref{klein}). Anyway, we are able to solve exactly
some important cosmological cases, then we can perfectly control 
these two different asymptotic behaviours. In particular, we can understand
how the asymptotic freedom depends upon initial data of the problem
(fine tuning).

In conclusion, we can say that, at least at a classical level, 
 asymptotic freedom seems to be a fundamental 
  feature also for gravity,
if the gravitational "constant" is supposed to be a function of a scalar 
field (and then of time).
A further step in our analysis is to see how the 
presence of ordinary matter
affects all the above considerations. Finally, another
important goal related to what we have done is to find the most
general conditions for a cosmological model to obtain asymptotic freedom.
That is, if we impose only that $G_{eff}\rightarrow 0$ for $t\ll 0$,
which are the cosmological models satisfying such a conditions?
These last topics are the subjects which we will try to understand
in a forthcoming paper.

\vspace{3. mm}

\begin{centerline}
{\bf REFERENCES}
\end{centerline}
\begin{enumerate}
\item\label{birrell}
N.D. Birrell and P.C.W. Davies, {\it Quantum Fields in Curved Space}
(Cambridge Univ. Press, Cambridge, 1986)
\item \label{guth}
A. Guth, \pr  {\bf D 23} (1981) 347\\
A. Albrecht and P.J. Steinhardt, \prl {\bf 48} (1982) 1220
\item\label{la}
C. Mathiazhagen and V.B. Johori, \cqg {\bf 1} (1984) L29\\
D. La  and P.J. Steinhardt,  {\it Phys. Rev. Lett.} {\bf 62} (1989) 376 \\
D. La,  P.J. Steinhardt  and E.W. Bertschinger, 
{\it Phys. Lett.} {\bf B 231} (1989) 231 
\item\label{matter}
S. Capozziello, R. de Ritis, and P. Scudellaro,  \ijmp {\bf D 2} (1993) 
463;
S. Capozziello and R. de Ritis, \pl {\bf A 195} (1994) 48
\item\label{quigg}
C. Quigg, 
{\it Gauge Theories of Strong, Weak...},
Addison--Wesley Publ. Co. (1983);
P.D.B. Collins, A.D. Martin and E.I. Squires,
  {\it Particle Physics and Cosmology}, 
John Wiley and Sons, New York (1991);
U. Amaldi et al., \pl {\bf B 260}, (1991) 447
\item\label{meyerhof}
W.E. Meyerhof, {\it Elements of Nuclear Physics}, Mc Graw--Hill, New York
1967
\item\label{appelquist}
H. Kendall, {\it Proc. Vth Int. Symp. Electron and Photon Interactions
at High Energies}, Cornell Univ. (1971)
\item\label{bjorken}
J.D. Bjorken, \pr {\bf 179} (1969) 1547
\item\label{markov}
M.A. Markov, {\it Phys. Uspekhi} {\bf 37} (1994) 57;\\
O. Bertolami, J.M. Mour\~ao and J. P\'erez--Mercader,
\pl {\bf B 311} (1993) 27
\item\label{quartic}
S. Capozziello, R. de Ritis, C. Rubano, P. Scudellaro,
to appear in \ijmp {\bf D} (1995)
\item\label{deser}
S. Deser, \pl {\bf B 134} (1984) 419;
Y. Hosotani, \pr {\bf D 32} (1985) 1949;
O. Bertolami \pl {\bf B 186} (1987) 161
\item\label{nohair}
S. Capozziello, R. de Ritis and P. Scudellaro, \pl {\bf A 188} (1994) 130
\item \label{nmc}
S. Capozziello and R. de Ritis, \pl {\bf A 177} (1993) 1;\\
S. Capozziello and R. de Ritis, \cqg {\bf 11} (1994) 107;\\
S. Capozziello and R. de Ritis, \pl {\bf A 203} (1995) 283.
\item\label{stringhe}
S. Capozziello, R. de Ritis, and C. Rubano \pl {\bf A 177} (1993) 8;\\
S. Capozziello and R. de Ritis \ijmp {\bf 2D} (1993) 373
\end{enumerate}
\vfill

\end{document}